\documentclass{aa}
\usepackage{amssymb}
\usepackage{times}
\usepackage{graphics}
\usepackage{psfig}

\def\asec{\ifmmode ^{\prime\prime}\else$^{\prime\prime}$\fi}
\def\lea{\ifmmode ^{<}_{\sim} \else $^{^{<}_{\sim}}$\fi}
\def\gea{\ifmmode ^{>}_{\sim} \else $^{^{>}_{\sim}}$\fi}
 
\begin{document}
\thesaurus{06.                  
		(08.14.1;       
		08.16.7;        
		03.20.4;        
		03.20.1)        
		}

\title{\bf $BVR_cI_c$  light curves of GRB970508 optical remnant and colors
 of underlying host galaxy.}
\author{S.V. Zharikov\inst{1},
V.V. Sokolov\inst{1},
Yu. V. Baryshev\inst{2}}
\offprints{S. Zharikov:zhar@sao.ru }

\institute{Special Astrophysical Observatory of R..A.S,
 Karachai-Cherkessia, Nizhnij Arkhyz, 357147 Russia; zhar,sokolov@sao.ru
	  \and
Astronomical Institute of St.Petersburg University, St.Petersburg
198904, Russia; yuba@aispbu.spb.su}

 \date{Received / Accepted }

\authorrunning{Zharikov S.V. et al.}
 \titlerunning{$BVR_cI_c$  light curves of GRB970508 optical remnant.}

\maketitle

\begin{abstract}
Optical observations of the GRB970508 optical remnant were continued
with the 6-m telescope of SAO RAS in standard the $BVR_cI_c$ bands
in Oct.-Dec. 1997 and in Jan. 1998.
The results of  photometry of the GRB970508 remnant and
of three nearby galaxies are presented.
The $BVR_cI_c$ light curves of the GRB970508 remnant can be described by a
power law plus a
constant $F(t) = F_0 (t-t_0)^{\alpha} + C$.
In  determination of parameters   of the faint host galaxy
we used the results of our $BVR_cI_c$ photometry of May-August, 1997,
the data of recent observations with Keck-II
and WHT telescopes
and also the data of photometry of $R_c$ and $B$ bands obtained
by other authors based on our secondary standards.
The level-off from the initial power law decline seen
in the first months after the burst was observed in all bands.
The effect is the  strongest in the $I_c$ band: the difference on
the last date of the $I_c$ observations reaches $\sim$1.3 mags. The best
$\chi^2$-fits for $F_o$, $\alpha$, $C$ parameters of the data in each of 4 bands point to the 
presence of a
constant faint  source with
$I_c = 24.13\pm0.28$, $R_c = 25.55\pm0.19$, $V = 25.80\pm 0.14$,
$B = 26.68\pm 0.14$.
The average  $\alpha$  is $-1.23\pm0.04$.
The optical remnant
 has a power law spectrum with a spectral slope equal to $-1.10\pm0.08$
and does not change after the optical curve maximum.
The $BVR_cI_c$ spectrum together with
the absolute magnitude of the constant component
$M_{B_{rest}} = -17.5\pm0.3$,
and the linear size of the underlying host galaxy
$d \sim 3 \ kpc$
conform to a host galaxy,
such as a starburst dwarf,  a red starburst dwarf,
a irregular dwarf, a HII galaxy, or a blue compact dwarf galaxy.
All these types of dwarf galaxies show evidence
of starburst or post starburst activity. The galaxy G2
has a spectrum similar to that of the host GRB galaxy and lies at the
projected distance of $\sim 20$ kpc from GRB.
 
\end{abstract}

\section{Introduction.}
\label{intro}

After the launch of the BeppoSAX satellite (\cite{Boella})
 a new era began in the search and
optical investigation of gamma-ray bursts.
The precise locations determined with the Wide Field Cameras on board
the BeppoSAX satellite allowed one to discover X-ray
 (\cite{Costa}) and radio (\cite{Frail})
afterglow
of gamma-ray bursts.
\begin{figure}[t]
\centerline{
   \vbox{\psfig{figure=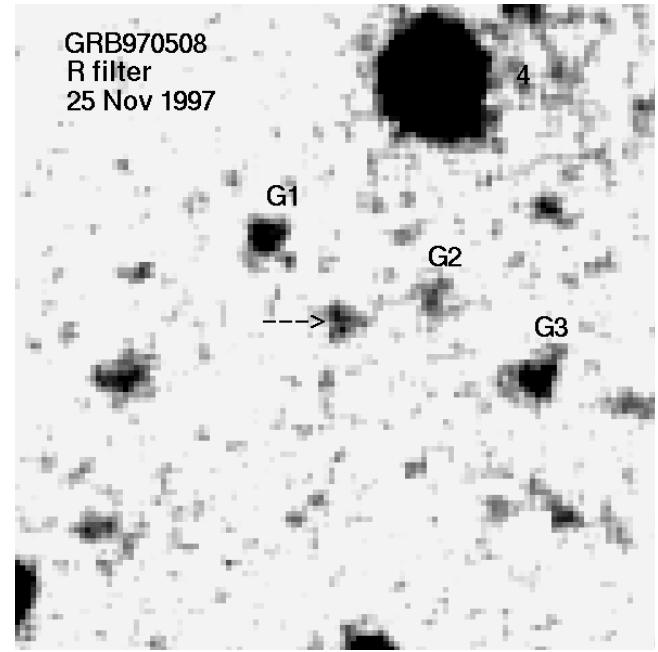,width=8.5cm,%
 bbllx=75pt,bblly=305pt,bburx=520pt,bbury=755pt,clip=}}}
\caption{ The $ R_c$ band field near GRB 970508 optical source.
 The image size is  $33^{\prime\prime}\times 33^{\prime\prime}$. N -top, E-right.
The G1, G2, G3 are nearby galaxies. The arrow denotes an optical remnant of
GRB970508.
}
\label{field}
\end{figure}
    The discovery of optical transients accompanying the gamma-ray bursts
was first announced by \cite{G1}. During a year
after the discovery of GRB970228, optical counterparts  were reliably
detected  for 4 other gamma-ray bursts: GRB970508 (\cite{Bond},
GRB971224 (\cite{Halpern}),
GRB980326 (\cite{G2}), GRB980329 (\cite{Dj2}).
The study of optical afterglow light curves 
and of the properties of underlying optical GRB counterparts 
gives a chance to understand the physics of the GRB phenomenon.
 
\begin{table*}[t]
\label{mag1}
\begin{center}
\caption{ The SAO RAS photometry of OT GRB970508 in Oct. 1997 - Jan. 1998.} 
\begin{tabular}{lrccccc} \hline
  UT & $ t-t_0$    &      $ B$   &  $ V$          &   $ R_c$      &      $ 
I_c$      &    \\ \hline
09.000 Oct.&154.12&               &               &  $24.30\pm0.2$ & 
                &    \\
09.000 Nov.&185.12&               &$25.10\pm 0.17$&  $24.7\pm0.15 $& 
                &    \\
25.00      &201.09&               &$25.50\pm 0.35$& 
               &$23.90\pm 0.14$  &    \\
25.00      &201.16&               &               &$24.7\pm 0.14$  & 
                &    \\
01.00  Dec$^*$.&206.01& $25.75\pm 0.30$&              &                & 
                &    \\
24.87  Jan.&  260.96    &                &              &$ 24.96\pm0.17$  &
                      \\
24.87      &  260.96    &                &$25.44\pm0.25$&               &
                 &  \\ \hline
\end{tabular}
\label{mag}
\end{center}
\end{table*}

\begin{table*}[t]
\label{magGal}
\begin{center}
\caption{ Magnitudes and colors of GRB970508 field objects
without galaxy extinction correction.}
\begin{tabular}{lrcccccc} \hline
Object &  $B$          &  $V$           &   $R_c$       & $I_c$         &
$B-V$        &  $V-R_c$    & $R_c-I_c$     \\ \hline
  G1   &$25.22\pm0.12$   & $24.61\pm0.07$  &$24.23\pm0.05$   &$24.21\pm0.16$      & $0.61\pm0.14  $  &$0.38\pm0.09$    &$0.02\pm0.17$  \\
  G2   &  $>26.7 $& $26.1\pm0.25$   &$25.31\pm0.12$   &$23.73\pm0.08$      & $>0.6$    &$0.79\pm0.25$    &$1.58\pm0.14$  \\
  G3   &$25.29\pm0.13$   & $24.79\pm0.08$  &$24.20\pm0.04$   &$23.99\pm0.09$      &$0.50\pm0.15$     &$0.59\pm0.09$    &$0.20\pm0.10$
\\ \hline
\end{tabular}
\label{magG}
\end{center}
\end{table*}

\begin{figure}[t]
\centerline{
   \vbox{\psfig{figure=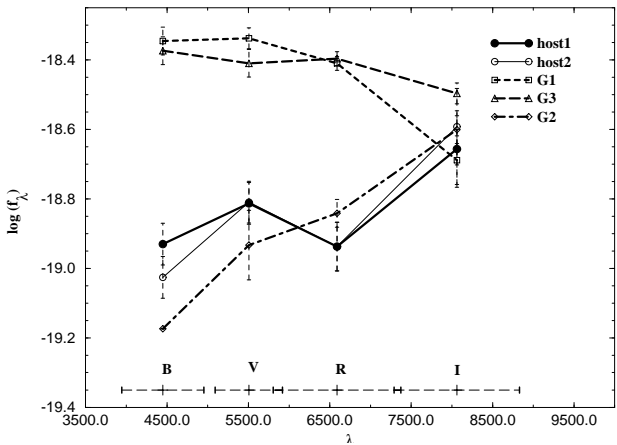,width=8.5cm,%
 bbllx=40pt,bblly=60pt,bburx=590pt,bbury=470pt,clip=}}}
\caption{Observational $BVR_cI_c$ spectra of GRB970508 remnant field 
objects (G1, G2 and G3) corresponding to Fig.1 are shown.
Two variants of broad-band spectra of the GRB host galaxy denoted
as "host1" and "host2" (see the text) are also shown.
The X axis is $\lambda (\AA)$, the Y axis is log($f_{\lambda}$).
The $f_{\lambda}$ is given in $ergs\ cm^{-2}\ s^{-1}\ \AA^{-1}$.
The FWHTs for the $BVR_cI_c$  bands are shown under the figures.
}
\label{specAll}
\end{figure}

This article presents the results of optical follow-up observations 
of the variable source associated with  the GRB970508 remnant.
The first results of our multi-color photometry
from May to  Aug. 1997 of the GRB970508 optical remnant were reported by
\cite{Sokolov}.
In this paper we report the results of a further $ BVR_cI_c $ photometry
performed until up to Jan. 1998.
The next section is dedicated to the results of CCD photometry of
the optical remnant in different filters and to an interpretation of 
the optical light curves with power-law $\chi^2$ fits. 
The photometry of nearby galaxies to the GRB970508 optical remnant
is given. Discussion and summary constitute Section 3.

\section{The field of GRB970508 optical counterpart.}
\label{section1}

Observations of the GRB970508 optical remnant were performed using
the primary focus CCD photometer of the 6m telescope SAO RAS (see
\cite{Sokolov}).
 The observations were carried out with the standard  (Johnson  -  Kron  -
Cousins) photometric $ BVR_{c}I_{c}$ system.
Here we present  the results of our observations in the period
Oct. 1997 - Jan. 1998.
The magnitudes in the $ BVR_{c}I_{c}$ bands
with associated errors for the GRB970508 optical counterpart
are given in Table.\ref{mag1}. Our last $B$ point (denoted by * in
Table.\ref{mag1}) was obtained by
A. Kopylov.
Fig.\ref{field} shows a part of the deep $R_c$ band image of
GRB980508 field.
An arrow points the GRB970508 optical remnant and nearby galaxies.
These galaxies are denoted as G1, G2 and G3 in Fig.\ref{field}.
 
The magnitudes and colors of the galaxies  G1, G2, G3  are given in
Table.\ref{magG}.
The photometry of those objects
was obtained using a digital sum of
all images of the GRB970508 field with a
seeing better than
$1.8^{\prime\prime}$. We made a photometric calibration using our secondary 
photometric standards from
\cite{Sokolov}. The total exposure times for each filter were
 $B$ 4800 sec., $V$ 14200 sec., $R_c$ 13950 sec., $I_c$ 10800 sec..
The galaxies G1, G3  are blue galaxies with close magnitudes and similar
spectra which probably attests that they have the same $z$.
The galaxy  G2 is red with $R_c-I_c=1.58\pm0.14$.
 
The magnitudes of $g = 24.5\pm0.4$ and $r=24.8\pm0.4$ for  G1 from 
\cite{Dj1} agree with
our V and $R_c$ magnitudes with subsequent offset
between broadband magnitudes in the Gunn photometric system and
the Johnson-Kron-Cousins photometric system
$ V = g + 0.031$ É $ R_c = r - 0.343$ (see. \cite{Frei}) and
$\approx0.6$ brighter than
the STIS/$R$ magnitude  ($R=24.8\pm0.2$) from \cite{Pian}.
The difference may be related to  the $R$ magnitude interpolated by
\cite{Pian} for G1. This interpolation is done by
a power-law spectrum consistent with the ($V-H$) color index.
For the G2 galaxy our $R_c =25.31\pm0.12$ magnitude agrees with the
\cite{Pian} ($25.5\pm0.2$) magnitude.
 
\begin{figure*}
\centerline{
   \vbox{\psfig{figure=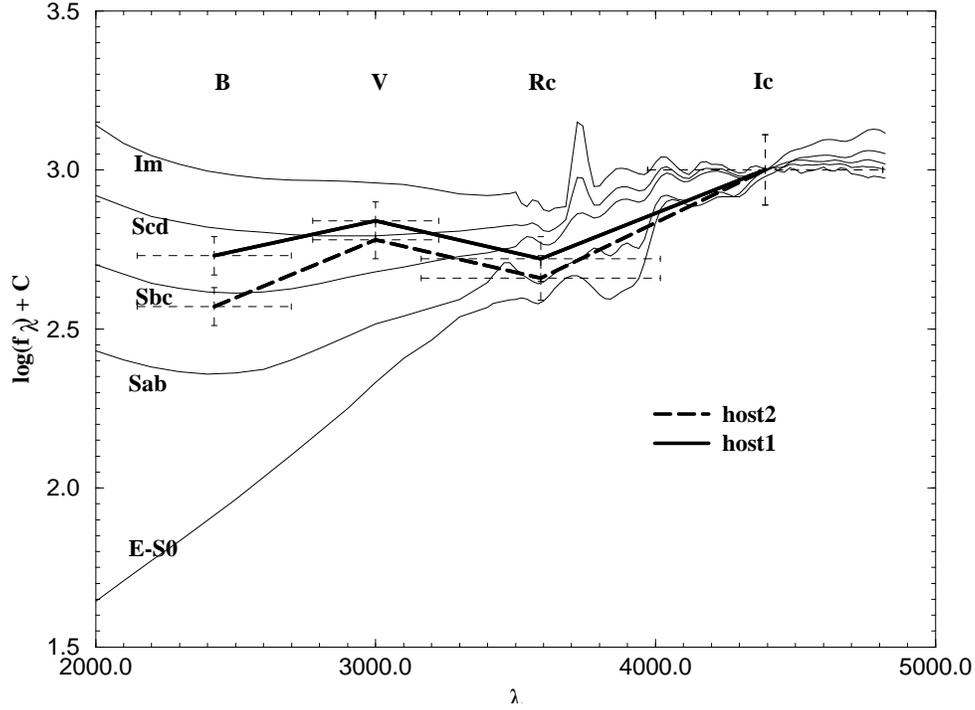,width=13cm,%
 bbllx=60pt,bblly=60pt,bburx=590pt,bbury=470pt,clip=}}}
\caption{
The  $BVR_cI_c$ broad-band spectrum $log(f_{\lambda})+Constant$
 for the host object
(with "host1" and "host2" fits) for $z=0.835$ is shown.
The observing bands have been shifted to the rest frame of GRB970508.
The spectral energy distribution for different morphological
 types of galaxies,
E-S0, Sab, Sbc, Scd, Im, from Pence  (1976)
are shown for comparison.
}
\label{spec2}
\end{figure*}

\begin{table*}[t]
\label{fit}
\begin{center}
\caption{ The $\chi^2$ fits for $BVR_cI_c$ lights curves of GRB 970508
remnant ("host1" case).}
\begin{tabular}{lccccc} \hline
Filter            & $\chi^2 /(d.o.f.)$ & $F_0$(mag)  & $\alpha$
       &  F (mag)       & $log(F (\frac{erg}{cm^2 s \AA}))$ \\ \hline
$B$   &  15.2/13                &
$19.617\pm0.15$& $-1.28\pm0.04$ & $26.68\pm0.14$ & $-18.88\pm0.06$
		   \\
$V$      &  35.1/13                &
$19.236\pm0.14$& $-1.22\pm0.04$ & $25.80\pm0.14$ & $-18.76\pm0.08$
		    \\
$R_c$  &  164/37                 &
$18.779\pm0.11$& $-1.23\pm0.02$ & $25.55\pm0.19$ & $-18.89\pm0.07$
		    \\
$I_c$   &  43.9/11                &
$18.330\pm0.27$& $-1.18\pm0.07$ & $24.13\pm0.28$ & $-18.61\pm0.11$
		    \\
\hline
\end{tabular}
\label{mag2}
\end{center}
\end{table*}
Observational broad-band spectra of nearby galaxies
are shown in Fig.\ref{specAll}.
 
\begin{table*}[t]
\label{fit1}
\begin{center}
\caption{ The $\chi^2$ fits for $BVR_cI_c$ light curves of GRB
970508 remnant with $\alpha=-1.23$ ("host2" case).}
\begin{tabular}{lcccc} \hline
Filter & $\chi^2/(d.o.f.)$  & $F_0$ (mag)     &  $F$ (mag)
& $log(F (\frac{erg}{cm^2 s \AA}))$                \\ \hline
$B$    & 17.9/14         & $19.668\pm0.15$ & $26.92\pm0.14$        & $ -18.98\pm0.08$    \\
$V$    & 34.7/14                 & $19.232\pm0.14$ & $25.79\pm0.14$& $-18.76\pm0.08$     \\
$R_c$  & 164/38                 & $18.779\pm0.11$ & $25.55\pm0.19$ & $-18.89\pm0.07$    \\
$I_c$   & 44.9/12               & $18.289\pm0.27$ & $23.97\pm0.28$ & $-18.54\pm0.11$    \\
        \hline
\end{tabular}
\label{fit2}
\end{center}
\end{table*}

\section{The GRB970508 $BVR_cI_c$ light curves.}
\label{section2}
\begin{figure*}
   \vbox{\psfig{figure=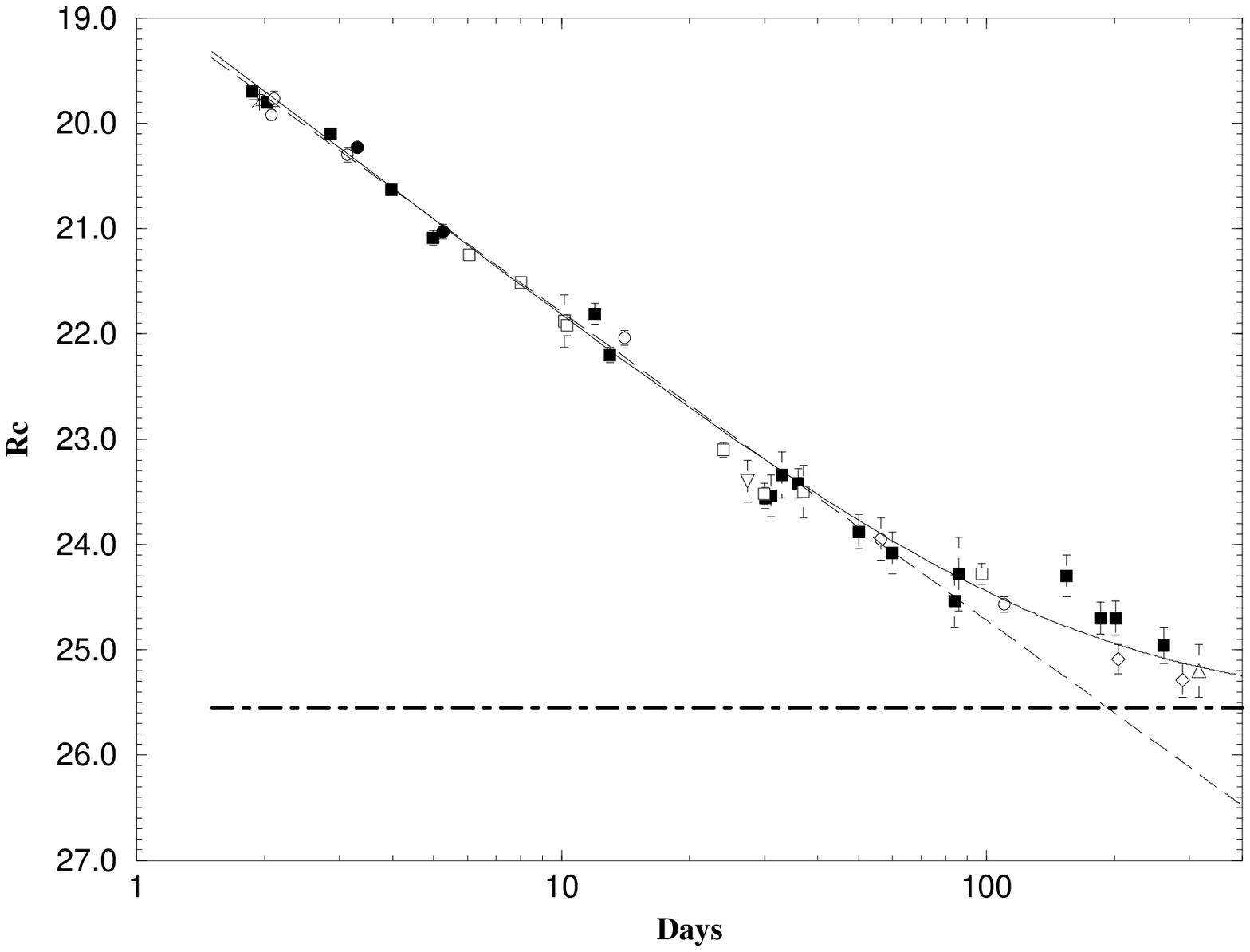,width=14cm,%
 bbllx=50pt,bblly=50pt,bburx=580pt,bbury=450pt,clip=}}
   \vbox{\psfig{figure=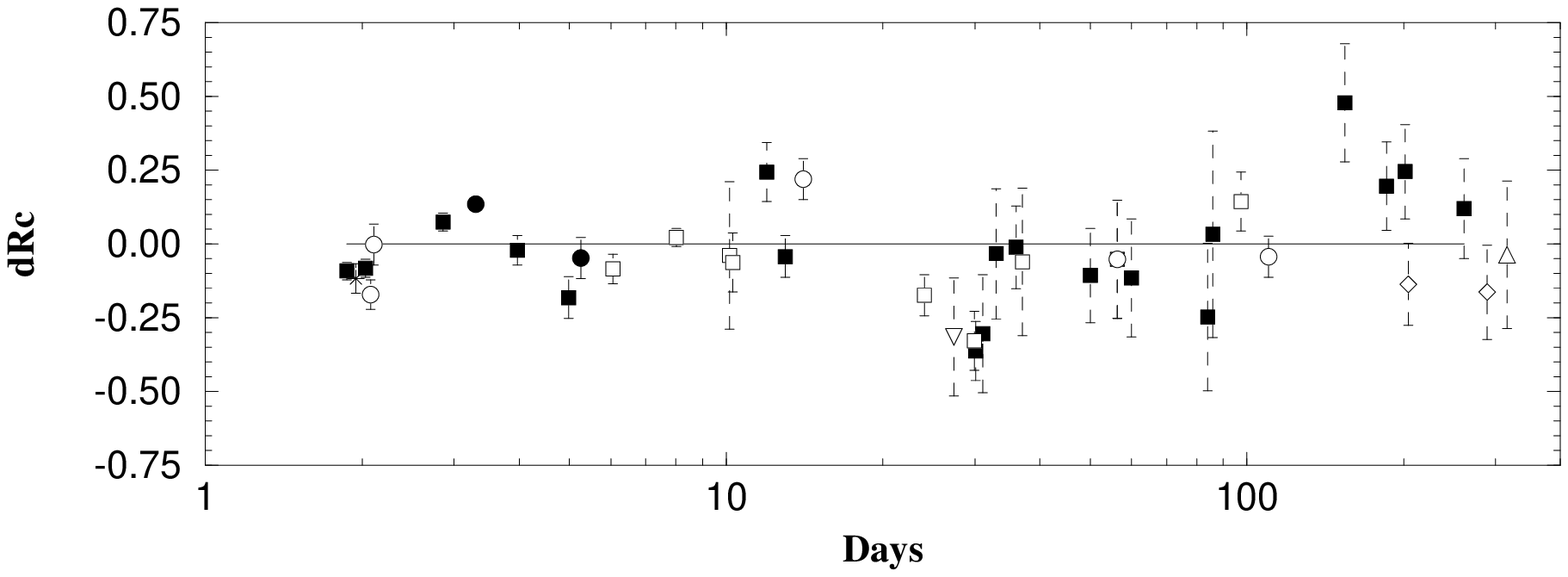,width=14cm,%
 bbllx=40pt,bblly=40pt,bburx=580pt,bbury=280pt,clip=}}
\caption{The $R_c$ light curve of the GRB970508 optical remnant. The
power law fit to the May-Aug. 1997
from Sokolov et al., 1998 is represented by the long-dashed  lines.
 Indicated by the solid line is a fit by the whole data of a power law plus
a constant with the parameters from Table 3.
The dot-dashed lines correspond
 to constant object magnitudes for "host1, host2 "  cases.
The lower panel: the fit to the power law  decay plus
a constant ("host1") subtracted from the observational data.
The symbols correspond to Sokolov et al. (1998, $\blacksquare$),
 Galama et al., (1998, $\circ$),
Garsia et al., (1998; $\bullet$), Mignole et al., (1997) (*), Metsger et al., (1997; $\bigtriangledown$),
 Pedersen et al., (1998, $\square$),
 Bloom et al., (1998, $\diamond$),
Castro-Tirado et al., (1998, $\triangle$).}
\label{f4}
\end{figure*}

\begin{figure*}
\centering{
   \vbox{\psfig{figure=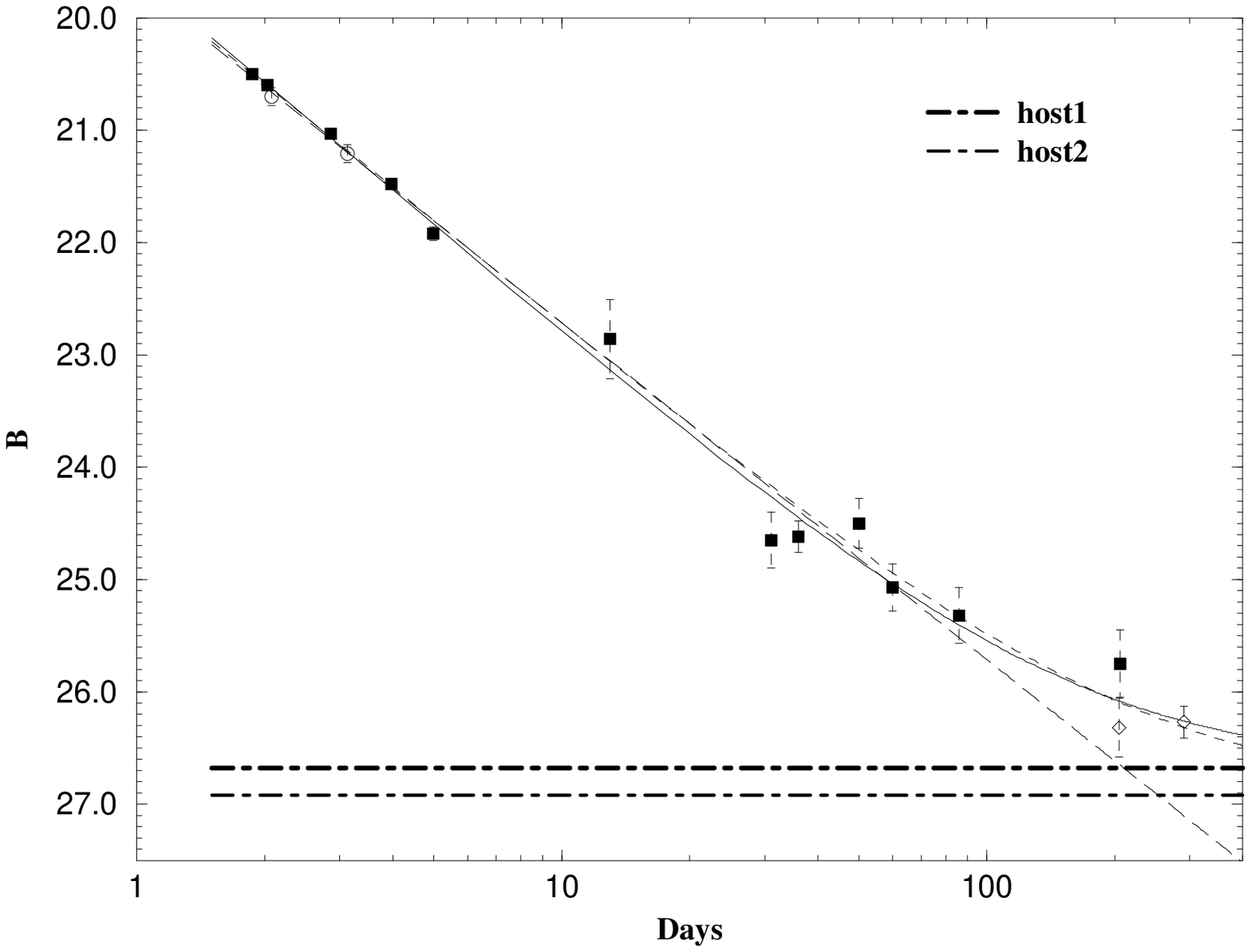,width=14cm,%
 bbllx=50pt,bblly=50pt,bburx=580pt,bbury=450pt,clip=}}
   \vbox{\psfig{figure=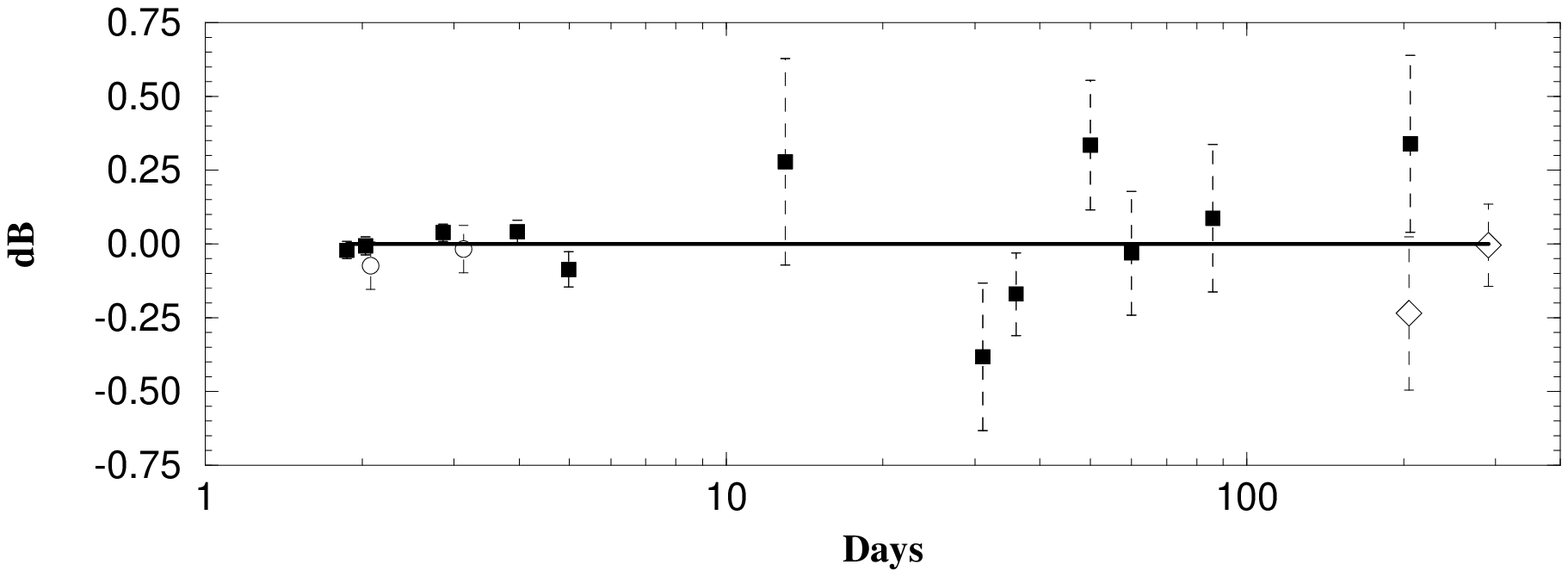,width=14cm,%
 bbllx=40pt,bblly=40pt,bburx=580pt,bbury=280pt,clip=}}
}
\caption{The $B$ light curve of the GRB970508 optical remnant. The
power law fit to the May-Aug. 1997
from Sokolov et al., 1998 is represented by the long-dashed  lines.
 Indicated by the solid line is a fit by the whole data of a power law
plus
a constant with parameters from Table 3.
 The power law fit with $\alpha=-1.23$
is given by the  dashed line. Dot-dashed lines correspond
 to constant object magnitudes for "host1" and "host2" case.
The lower panel: the fit to the power law fading plus
a constant ("host1") subtracted from the observational data.
The symbols correspond to Sokolov et al. (1998, $\blacksquare$),
 Galama et al., (1998, $\circ$),
 Bloom et al., (1998, $\diamond$).
}
\label{f5}
\end{figure*}

\begin{figure*}
\centering{
   \vbox{\psfig{figure=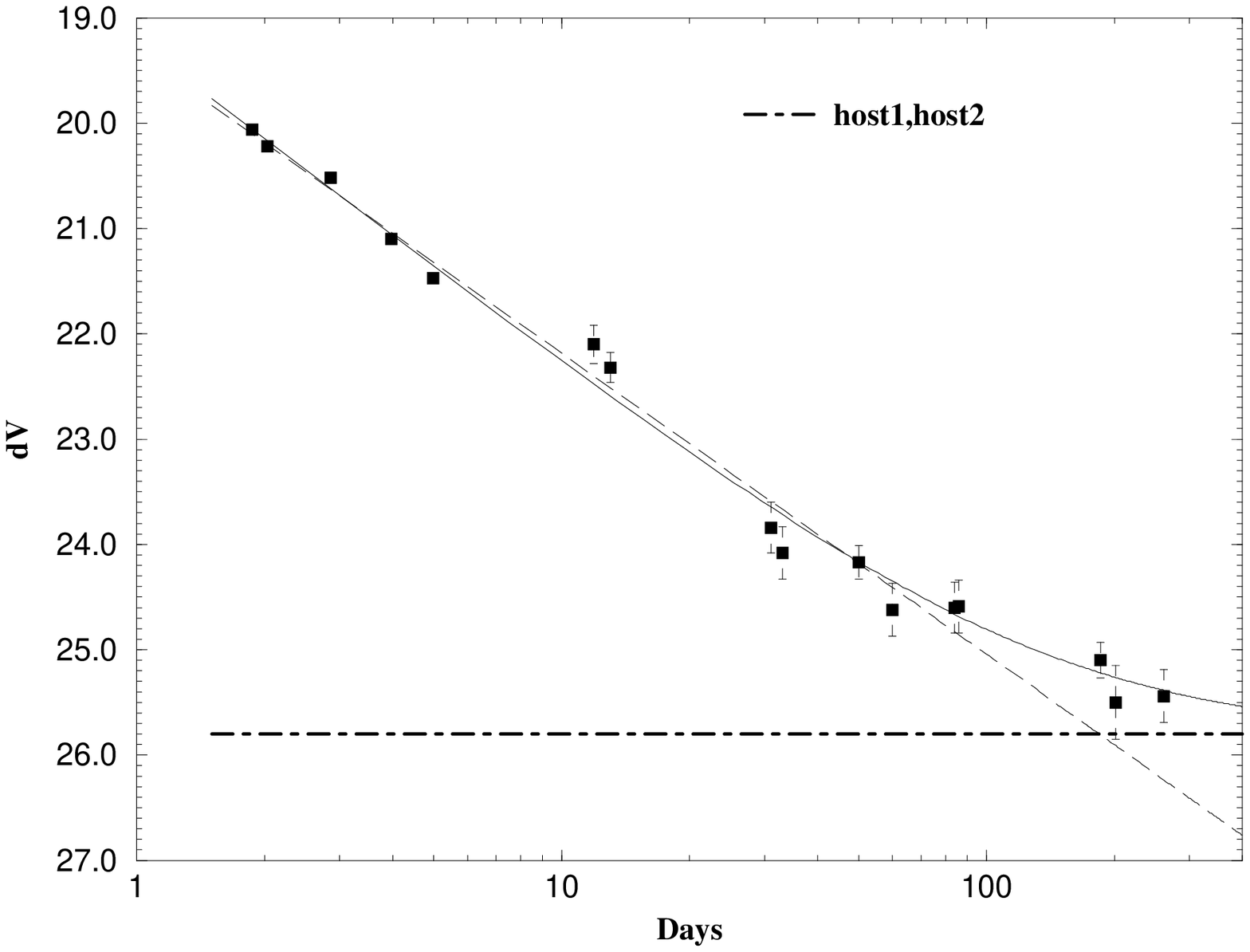,width=14cm,%
 bbllx=50pt,bblly=50pt,bburx=580pt,bbury=450pt,clip=}}
  \vbox{\psfig{figure=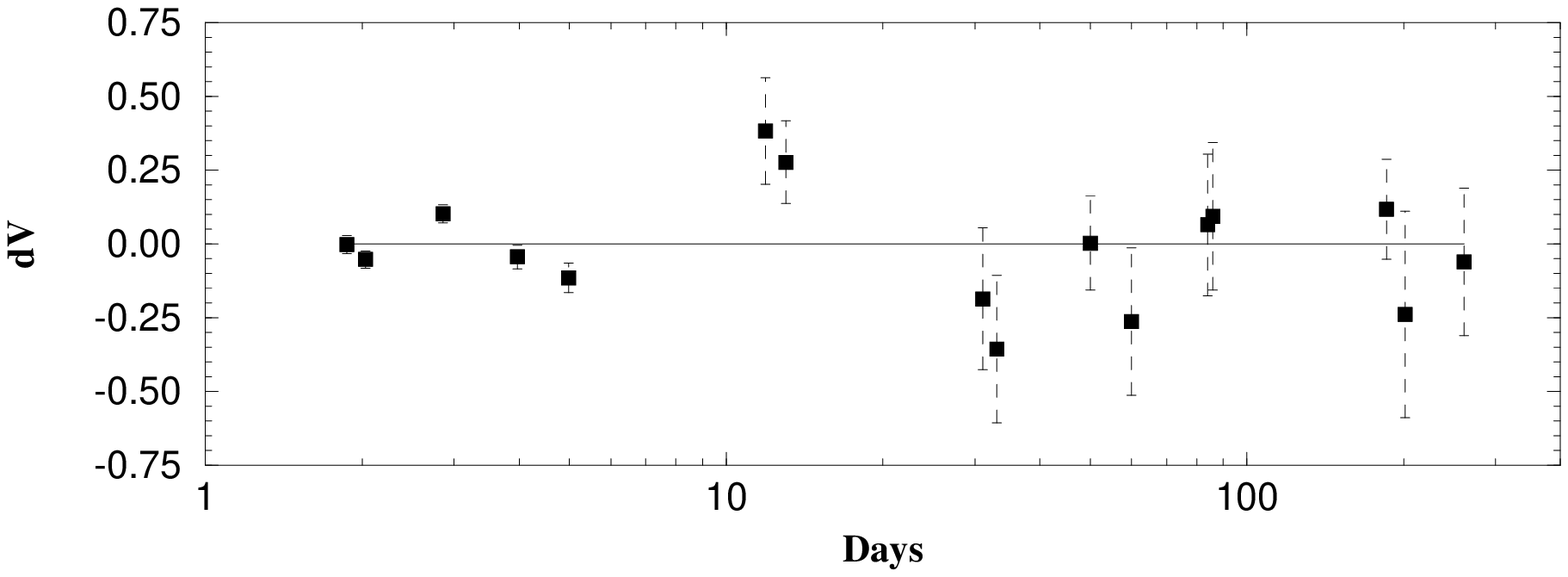,width=14cm,%
 bbllx=40pt,bblly=40pt,bburx=580pt,bbury=280pt,clip=}}
}
\caption{ The $V$ light curve of  the GRB970508 optical remnant.
the Lines and symbols are the same as in  Fig.4.
}
\label{f6}
\end{figure*}

\begin{figure*}
\centering{
   \vbox{\psfig{figure=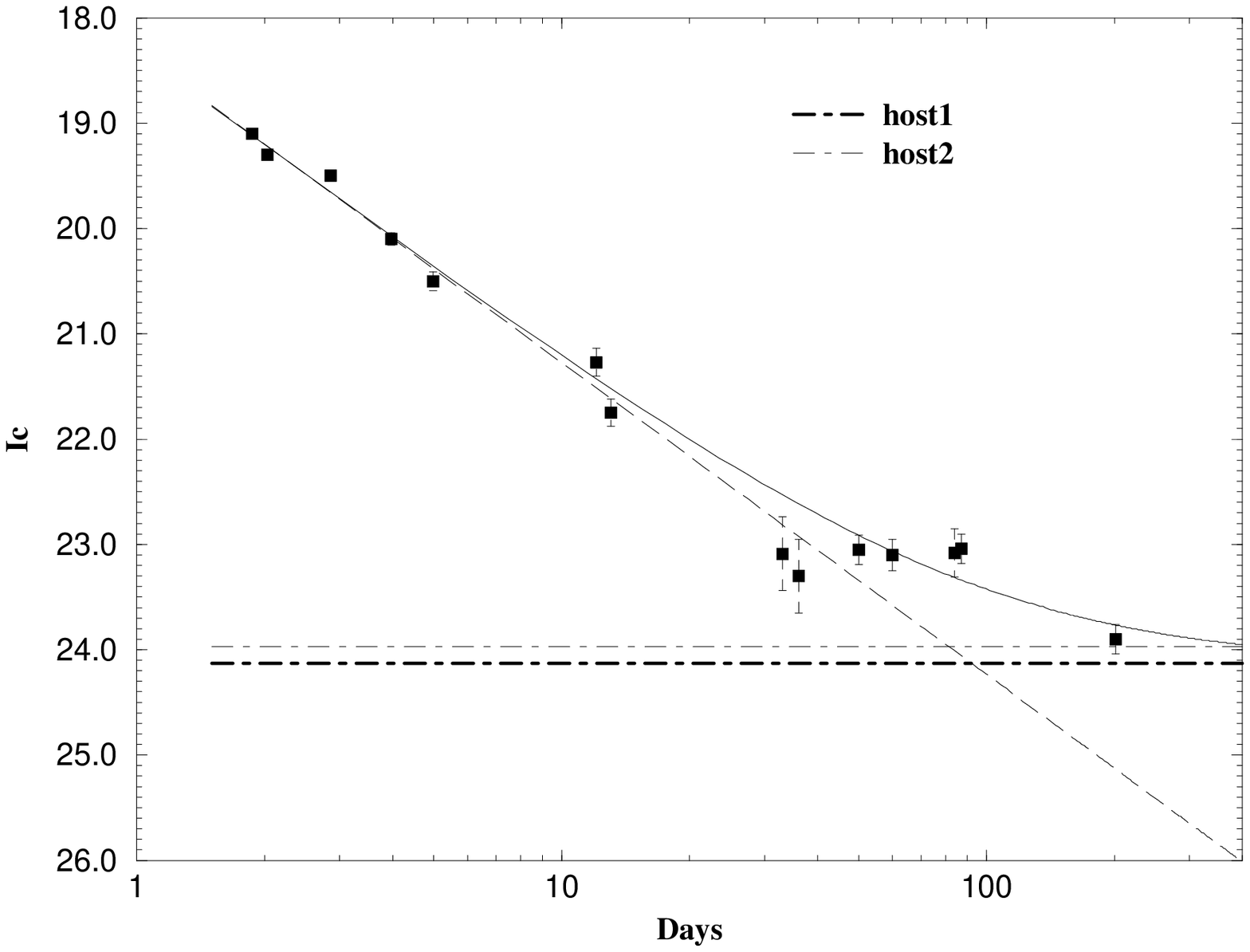,width=14cm,%
 bbllx=50pt,bblly=50pt,bburx=580pt,bbury=450pt,clip=}}
 \vbox{\psfig{figure=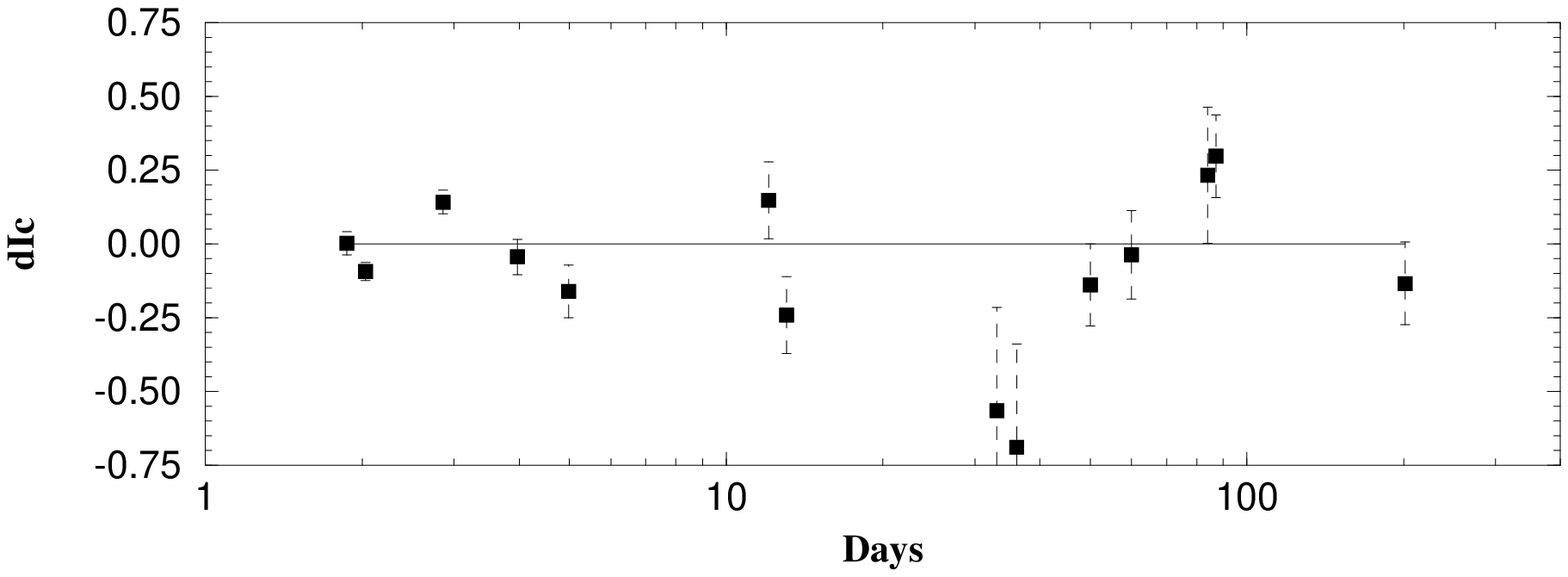,width=14cm,%
 bbllx=40pt,bblly=40pt,bburx=580pt,bbury=280pt,clip=}}
 }
\caption{ The $I_c$ light curve of  the GRB970508 optical remnant.
The lines and symbols are the same as in Fig.4.
}
\label{f7}
\end{figure*}

In Figs.\ref{f4},\ref{f5},\ref{f6},\ref{f7} we present the $BVR_cI_c$-band light curves of the optical
remnant of  GRB970508.
We included here the results of our photometry of May 1997- Jan 1998 
(labled by $\rule{2mm}{2mm}$),
the data of recent observations with the Keck-II
(\cite{Bloom}  $\diamond$) and the WHT telescopes (\cite{Castro}
$\triangle$)
 and  the data of $R_c$ and $B$ band photometry
obtained by other authors based on our secondary standards (\cite{Pedersen}
 $\square$;
\cite{Garsia}, $\bullet$;
\cite{Galama}, 1998, $\circ$ ). Also we have included
two points
from \cite{Mignoli} (*) and \cite{M1} ( $\bigtriangledown$).
 
\subsection{The "host1" case.}
\label{secsion2.1}

We  fitted the data by a power law plus a constant $F(t) = F_o t^{\alpha} 
+C$
in all bands using the $\chi^2$ fit (denoted as "host1"). The $F_o$ , $\alpha$
and $C$ are fit parameters independent for every band.
 Our fit results are presented in Table.3.
   We switched from  stellar  magnitudes  of  the
Cousins system to absolute fluxes in $erg\ cm^{-2}\ s^{-1}\ \AA^{-1}$ with
the data on $\alpha$Lyr  published  by  \cite{Fukugita}.
 
Magnitude errors were calculated with the uncertainty of the fit
$\sigma^2_{mag} = \sigma^2_{\chi^2}/\sqrt{N_{point}}$.
 
For a host object we found:
 $B=26.68\pm0.14$, $V=25.80\pm0.14$, $R_c=25.55\pm0.19$, $I_c=24.13\pm0.28$. 
Taking into account the galactic absorption $  E(B-V)=0.03  $  for color
indexes we have: 
$$(B-V)=0.88\pm0.20; \ (B-V)_0=0.85\pm0.20$$
$$ (V-R_c)=0.25\pm0.24; \  (V-R_c)_0=0.23\pm0.24$$
$$ (R_c-I_c) =1.42\pm0.34; \ (R_c-I_c)_0 =1.40\pm0.34$$
    The optical fluxes for the GRB970508 remnant in 300 day after the gamma-ray
burst
is $\sim$19\% in $B$ , $\sim24$\% in $V$ , 
$\sim23$\% in $R_c$ , $\sim25$\% in $I_c$ 
relative to  the host object fluxes in the $B, V, R_c, I_c$ bands respectively.
The decay slope $\alpha$ (for the decay phase) of optical
brightness of the GRB970508 remnant
is constant, $<\alpha>=-1.23\pm0.04$, within the errors for the 4 bands
(see Table.3).
It follows that the optical remnant spectrum does not change after the optical
light curve maximum and the GRB970508 optical afterglow has a power law
spectrum with a spectral slope
equal to $-1.10\pm0.08$ (\cite{Sokolov}, \cite{M2}).

\subsection{The "host2" case.}
\label{secsion2.2}
 We also fitted our data by
 a power law plus a constant $F(t) = F_o t^{\alpha} +C$ with $\alpha = 
-1.23$ (denoted as "host2").
We present results of this fit in Table.\ref{fit2}.
Colors  for a constant object from the last
fit are
$$(B-V) =1.12\pm0.20; \ (B-V)_0=1.10\pm0.20$$
$$(V-R_c)=0.26\pm0.24; \ (V-R_c)_0=0.24\pm0.24$$
$$(R_c-I_c)= 1.47\pm0.34; (R_c-I_c)_0 =1.45\pm0.34,$$
without and with $E(B-V)=0.03$ correspondingly.

The spectra of the optical transient obtained with the Keck-II 10-m telescope
on 11 May
near the brightness maximum demonstrated several absorption features which 
were interpreted
as MgII and Fe II with $z=0.835$ and Mg II with $z=0.767$ (\cite{M1}).
The $BVR_cI_c$ broad-band spectra for the "host1" and "host2" cases in the rest
frame of the host object for $z=0.835$ are displayed in Fig.\ref{spec2}.
The spectral energy distribution for galaxies with different morphological
types, E-S0, Sab, Sbc, Scd, Im,
from \cite{Pence} are shown for comparison.
 
The host object color $R_c-I_c$ corresponds to  E-Sab galaxy types.
The color indexes $B-V$ and $V-R$ show flat spectrum which conforms
with Sbc-Im galaxies.

\section{Discussion.}
\label{section3}
The light curves of the  GRB970508 optical remnant  show deviation from
single power law decay for the optical transient.
On the average the brightness fading in all filters can be described by a power law with
an additional constant source. 
However we note that the introduction of a constant to the power law fit did
not improve the $\chi^2$ substantially.
From too large  $\chi^2$ we conclude that a single power law plus
a constant is not a
good description of all observational data.
A single power law describes the light curve on the average on a large-time scale of
brightness decay.
But besides the exponential law of the source brightness fading on
first days after the maximum (\cite{Sokolov})
a further  intrinsic variability of the fading is possible too.
 
Observations in four optical bands during all brightness fading period of the  GRB970508 optical
counterpart allowed us to make the following conclusions on the character of optical
behavior of the GRB970508 remnant and the properties of its host galaxy.

1)
In the $I_c$  band the decay slows down from  the 31th day onwards.
 We conclude that the evidence for  flattening of the light curve at
late times is convincing.
The brightness of the GRB optical remnant was $\sim25$\% from the host galaxy brightness
in 300th day from the gamma-ray burst.
We note here that HST/NICMOS flux measurements (\cite{Pian})
in the $ H $ band  gave
$6.2\pm1.5$ $\mu$Jy, but  not $3\pm1$ $\mu$Jy as was expected  for
the power-law brightness fading in the ground-based $ I $ and $ K $  bands.
It was necessary to extrapolate these data to the $ H  $  band,
assuming the same power-law spectrum slope which had been  observed at the
brightness maximum phase
(\cite{M2}; \cite{Sokolov}).

2) The temporal slope $\alpha$ of the  decay phase of optical
brightness of the GRB970508 remnant
is constant, $<\alpha>=-1.23\pm0.04$, within the errors over all 4 bands.
It follows that the optical remnant spectrum does not change after the optical
light curve maximum and the GRB970508 optical afterglow has the power law
spectrum
with the spectral slope
equal to $-1.10\pm0.08$.

3) The $R_c = 25.55\pm0.19$ 
magnitude for the possible underlying galaxy is consistent with the limits
derived by \cite{Pian} ( $R_c>25.5$) and
the host object R magnitude is consistent from \cite{Pedersen} ($R_c=25.5\pm0.4$).
The results of our power law fits with
the additional constant source is in good
 agreement with the late observations in the  $B$ and the $R_c$ bands
with the Keck-II (\cite{Bloom}) and the WHT (\cite{Castro}).

4) If the observed $BVR_cI_c$ spectrum belongs to the host galaxy of
GRB, then we can calculate some physical parameters of the
galaxy and its possible type.
For Friedmann model with $H_0 = 75\ km\ s^{-1} \ Mpc^{-1}$, $q_0 = 0.5$, and,
assuming the redshift of the host galaxy $z = 0.835$, we obtain an
absolute magnitude $M_{B_{rest}} = -17.5 \pm0.3$ in the rest frame
of the host galaxy. It should be noted that the effective wavelength of the $R_c$
$I_c$ bands just corresponds to the rest frame $U_{rest}$ and $B_{rest}$.
This allows us
to calculate the K-correction, which is $+1.3\pm0.3$ and the color
$(U - B)_{rest} = +0.3 \pm0.3$.
We estimate the linear size of the host galaxy according to the HST/STIS
observations (Pian at al. 1998). If it is $R_c = 25.55$, then the upper
limit of the size of a possible extended source may be a few tenths
of an arcsecond. From the  HST/NICMOS observations an additional limit ($0^{\prime\prime}.4$)
is obtained for underlying extended component. With the understanding 
that the true size may  be slightly larger or smaller, we use for the effective
linear size the value of $\sim 3$ kpc.
Taking into account the spectrum in Fig.\ref{spec2}, we conclude that
the host galaxy may be either a starburst dwarf, or a red starburst dwarf,
or an irregular dwarf HII galaxy, or a blue compact dwarf galaxy.
All these types of dwarf galaxies show evidence of starburst
or post starburst activity.

5) The nearest faint galaxy G2 has a similar
spectrum and lies at a projection distance of about 20 kpc from
the GRB host galaxy and is possibly a member of a compact group
of galaxies. It is also possible that there is a galaxy on the
line of sight, which has $z = 0.767$ (\cite{M1}) and may
have an additional effect  on the spectrum (\cite{Natarajan}).

However it is possible that the $ BVR_cI_c$ spectrum within
the errors can be
fitted by a single power law
 $F_{\nu} \sim \nu^{-\beta}$ with $\beta \approx 3$.
 
 \acknowledgements{
 To the 6-m Telescope Program Committee for allocation of observational time
for  of the GRB identification program;
to A. Kopylov and
Director
of SAO Yu.Yu.Balega for help in our program.
The work was carried out under support of the
 "Astronomy" Foundation (grant 97/1.2.6.4), INTAS N96-0315 and RBFI N98-02-16542.
}

\end{document}